\documentclass[fleqn,10pt]{wlscirep}
\usepackage[utf8]{inputenc}
\usepackage[T1]{fontenc}
\usepackage[defaultcolor=orange]{changes}
\title{Turbulent convection as a significant hidden provider of magnetic helicity in solar eruptions}

\author[1,*]{Shin Toriumi}
\author[2,3]{Hideyuki Hotta}
\author[3]{Kanya Kusano}
\affil[1]{Institute of Space and Astronautical Science, Japan Aerospace Exploration Agency, 3-1-1 Yoshinodai, Chuo-ku, Sagamihara, Kanagawa 252-5210, Japan}
\affil[2]{Department of Physics, Graduate School of Science, Chiba University, 1-33 Yayoi-cho, Inage-ku, Chiba 263-8522, Japan}
\affil[3]{Institute for Space-Earth Environmental Research, Nagoya University, Nagoya 464‐8601, Japan}

\affil[*]{toriumi.shin@jaxa.jp}

\begin{abstract}
Solar flares and coronal mass ejections, the primary space weather disturbances affecting the entire heliosphere and near-Earth environment, mainly emanate from sunspot regions harbouring high degrees of magnetic twist. However, it is not clear how magnetic helicity, the quantity for measuring the magnetic twist, is supplied to the upper solar atmosphere via the emergence of magnetic flux from the turbulent convection zone. Here, we report state-of-the-art numerical simulations of magnetic flux emergence from the deep convection zone. By controlling the twist of emerging flux, we find that with the support of convective upflow, the untwisted emerging flux can reach the solar surface without collapsing, in contrast to previous theoretical predictions, and eventually create sunspots. Because of the turbulent twisting of magnetic flux, the produced sunspots exhibit rotation and inject magnetic helicity into the upper atmosphere, amounting to a substantial fraction of injected helicity in the twisted cases that is sufficient to produce flare eruptions. This result indicates that the turbulent convection is responsible for supplying a non-negligible amount of magnetic helicity and potentially contributes to solar flares.
\end{abstract}
\begin{document}

\flushbottom
\maketitle

\section*{Introduction}

Solar flares and coronal mass ejections are the primary sources of space weather disturbances driving various plasma processes in the entire interplanetary space, including the near-Earth environment\cite{2011LRSP....8....6S,2011LRSP....8....1C,2011SSRv..159...19F}. The strongest events among these eruptions emanate from the solar active regions, in which the rotation and shear motions of strongly magnetised sunspots are often observed\cite{2019LRSP...16....3T}. Once a solar flare occurs and a coronal mass ejection is launched, a helical magnetic flux rope is observed in the interplanetary space\cite{2006SSRv..123...31Z,2007AnGeo..25.2453M}, which is the clearest evidence that solar flares are the sudden release of the excessive magnetic energy accumulated in the helical magnetic structure in the solar corona through magnetic reconnection and plasma instability\cite{2002A&ARv..10..313P,2006JGRA..11112103G,2012ApJ...760...31K}.

Magnetic helicity is a measure to quantify the topology of a magnetic field, such as twists, kinks, and internal linkages:
\begin{eqnarray}
H=\int \mbox{\boldmath $A$}\cdot\mbox{\boldmath $B$}\, dV,
\end{eqnarray}
where $\mbox{\boldmath $A$}$ is the vector potential of a magnetic field $\mbox{\boldmath $B$}$, i.e., $\mbox{\boldmath $B$}=\nabla\times\mbox{\boldmath $A$}$.  It is well conserved even in resistive magnetohydrodynamic (MHD) processes but gauge-invariant only if the magnetic field is fully contained in a closed volume.
Otherwise, the relative magnetic helicity,\cite{1984JFM...147..133B,1985CoPPC...9..111F}
\begin{eqnarray}
H_{\rm R}=\int (\mbox{\boldmath $A$}+\mbox{\boldmath $A$}_{\rm p})\cdot(\mbox{\boldmath $B$}-\mbox{\boldmath $B$}_{\rm p})\, dV.
\label{eq:hr}
\end{eqnarray}
is widely used because it is gauge invariant in any case, and the potential magnetic field $\mbox{\boldmath $B$}_{\rm p}$ is often adopted as a reference.
The total amount of injected helicity, $H_{\rm R}$, and the helicity injection rate (helicity flux),
\begin{eqnarray}
\frac{dH_{\rm R}}{dt}=2\int_{S}\left[
(\mbox{\boldmath $A$}_{\rm p}\cdot\mbox{\boldmath $B$}_{\rm h})V_{z}
-(\mbox{\boldmath $A$}_{\rm p}\cdot\mbox{\boldmath $V$}_{\rm h})B_{z}
\right]\, dS,
\label{eq:dhdt}
\end{eqnarray}
where the subscripts h and $z$ denote the horizontal and vertical directions, respectively, are widely used in analyses of flare-productive active regions. For instance, observations show that active regions with a larger amount of magnetic helicity tend to produce stronger flares\cite{2004ApJ...616L.175N,2007ApJ...671..955L,2012ApJ...759L...4T} and that the activity level is enhanced in association with the total injected helicity or the temporal variation of helicity flux\cite{2002ApJ...574.1066M,2002ApJ...577..501K,2008ApJ...686.1397P,2012ApJ...750...48P,2012ApJ...752L...9J}. This is why magnetic helicity attracts broad attention of the solar and heliophysics community in the context of flare prediction and forecasting.

Based on a copious amount of observational evidence, it is widely believed that the injection of magnetic helicity into the corona is mainly due to the emergence of twisted magnetic flux from the convection zone and the consequential motions of sunspots, such as rotation and shearing\cite{1987SoPh..113..267Z,1991SoPh..136..133T,1996ApJ...462..547L,2000ApJ...544..540L,2001ApJ...560L..95C,2002ApJ...577..501K,2003SoPh..216...79B,2017ApJ...834...56T,2022arXiv220406010T}. Therefore, investigating how magnetic helicity is accumulated in the corona as the magnetic flux emerges and builds up active regions is an important factor in understanding the occurrence of solar flares. However, it has almost never been considered how and to what extent the background convection affects the helicity injection, which is probably because we cannot probe the solar interior using direct optical observations. The detection of flux emergence in the convection zone using helioseismology is a promising technique but still in the development stage.\cite{2011Sci...333..993I,2013ApJ...762..131B,2013ApJ...770L..11T}

To solve this problem, we perform numerical simulations in which a twist-free magnetic flux tube emerges from the turbulent convection zone and examine whether the helicity injection into the upper atmosphere is negligibly small or comparable to the observations. Previous theoretical studies suggested that an untwisted emerging flux cannot reach the photosphere because it experiences counteracting aerodynamic drag (see\cite{2021LRSP...18....5F} and the references therein). Under the ideal condition with no background convection, even if the non-twisted flux tube reaches the photosphere, the helicity injection will be zero (as far as the emerging bipole keeps its geometrical symmetry). In this study, we explore the occurrence of helicity injection by calculating the case with convection and, if so, to what extent.

We perform our computations using radiative MHD code {\it R2D2}\cite{2019SciA....5.2307H} to reproduce the realistic turbulent thermal convection in the Sun. We use the convection model that is in a statistically steady state as the initial condition ($t=0$). We do not consider solar rotation in our model; therefore, the net kinetic helicity of the background flow field is negligibly small. The simulations with no solar rotation and thus infinitesimal net kinetic helicity enable investigation of the magnetic helicity injection that is caused purely by turbulence.

At $t=0$, a magnetic flux tube with an axial field strength of 12 kG and a typical radius of 8 Mm is placed at a depth of 22 Mm in the rectangular computational box that covers the entire solar convection zone, with the box spanning over $0\leq x\leq 98.3 \ {\rm Mm}$, $0\leq y\leq 98.3\ {\rm Mm}$, and $-201\ {\rm Mm}\leq z\leq 676\ {\rm km}$ (bottom panel of Figure \ref{fig:tile}). The bottom boundary ($-201\ {\rm Mm}$) is deeper than most previous convective flux emergence simulations, which were at most $-30\ {\rm Mm}$\cite{2010ApJ...720..233C,2014ApJ...785...90R,2017ApJ...846..149C}. Thus, the present model allows us to investigate the effects of large-scale convection on the emergence of magnetic flux and the resultant sunspot formation\cite{2019ApJ...886L..21T,2020MNRAS.494.2523H,2020MNRAS.498.2925H,2022MNRAS.517.2775K}. In the initial state, a mechanical balance between the flux tube and surroundings is achieved by lowering the entropy inside the tube. Therefore, the magnetic flux starts rising in response to the background velocity field without artificial buoyancy. For the purpose of comparison, we calculate two additional cases where the flux tubes are weakly and strongly twisted (right-handed twist), but the background convection field remains the same. The twist strengths of the weakly- and strongly-twisted tubes are $q_{\rm cr}/4$ and $q_{\rm cr}/2$, respectively, where $q_{\rm cr}\,(=0.125\ {\rm Mm}^{-1})$ is the critical twist for the kink instability\cite{1996ApJ...469..954L}, namely, these tubes are stable against the instability. In all three cases, the total magnetic flux in the axial direction is on the order of $2\times 10^{22}\ {\rm Mx}$.

\section*{Results}

\subsection*{General evolution}

\begin{figure}[ht]
\centering
\includegraphics[width=0.9\linewidth]{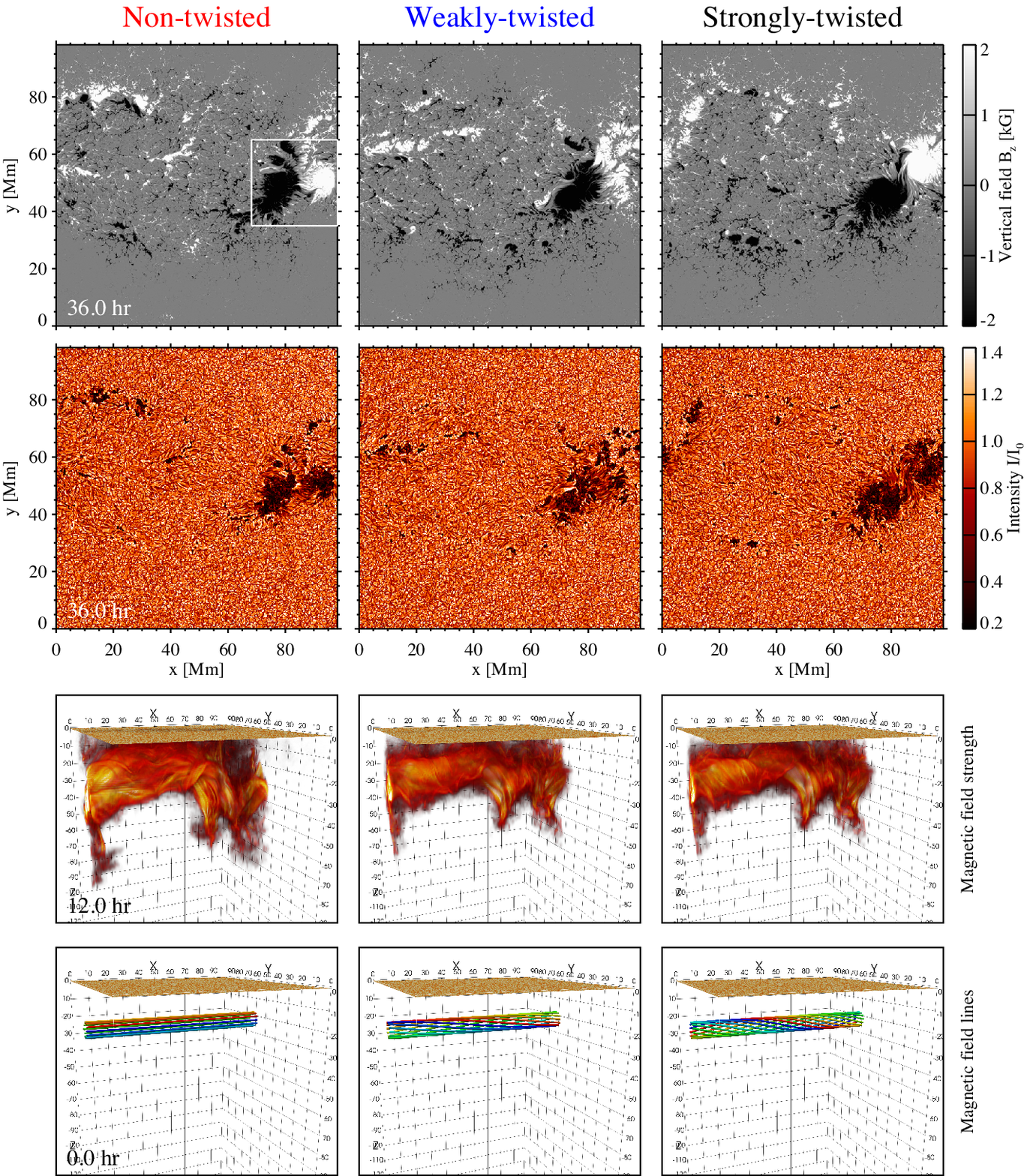}
\caption{The vertical magnetic field strength ($B_{z}$) at the photospheric surface, emergent intensity normalised by the average quiet-Sun intensity ($I/I_{0}$), three-dimensional (3D) volume rendering of the total magnetic field ($|\mbox{\boldmath $B$}|$) and initial magnetic field lines for the non-twisted, weakly-twisted and strongly-twisted cases. The white square emphasises the region where the two sunspots collide and rotate. The square also corresponds to the field of view of Figure \ref{fig:vortex}. A movie for the vertical field and intensity is available in the online Supplementary Information.}
\label{fig:tile}
\end{figure}

Figure \ref{fig:tile} shows the vertical magnetic field $B_{z}$ in the photospheric surface, the emergent intensity and the magnetic field strength and field lines in the 3D space for the three runs, i.e., the non-, weakly- and strongly-twisted cases. In this study, we define the layer where the optical depth is unity ($\tau=1$) as the photospheric surface and measure physical quantities to make direct comparisons with observations. Figure \ref{fig:tile} shows that the flux tube is levitated by a convective upflow located in the centre of the computational box in all three cases. Even the non-twisted flux tube successfully reaches the photosphere, in contrast to the previous prediction that a flux tube needs some twisting to maintain its cohesion against  aerodynamic drag\cite{2021LRSP...18....5F}.

From $t=20\ {\rm hr}$, the positive and negative magnetic elements distributed in the photosphere gradually assemble and build up sunspots with positive and negative polarities. The positive sunspot moves beyond the side periodic boundary and at around $t=30\ {\rm hr}$, the two sunspots collide with each other, eventually creating a strongly-packed bipolar sunspot ($\delta$-sunspot), which is known to be highly flare-active \cite{2000ApJ...540..583S}. This situation agrees with the scenario where a single flux system emerges at multiple locations to build up a colliding bipolar sunspot \cite{2014SoPh..289.3351T,2017ApJ...850...39T,2019ApJ...886L..21T}. The magnetic energy of the flux tube at the initial state is set to be the same for the three cases. However, once the sunspots are established in the photosphere, their decay is more rapid for the weaker twist cases because the local convection cells can more easily intrude into the sunspots and break them into pieces. In the twist-free case, the sunspots disappear by around $t=60\ {\rm hr}$, i.e., the lifetime in the photosphere is approximately 40 hr.

One remarkable feature of the developed sunspots is their continued rotation. In the strong twist case, the two sunspots rotate in the same clockwise direction because the flux tube, endowed initially with right-handed torsion, releases its twist as it appears in the photosphere\cite{2009ApJ...697.1529F,2015A&A...582A..76S,2019ApJ...886L..21T}. Observations show that many more flaring active regions show rotations of bipolar sunspots in the same directions than in the opposite directions\cite{2008MNRAS.391.1887Y}. Of particular note, however, is that the sunspots in the no-twist case also exhibit rotations, with the negative sunspot in a clockwise direction and the positive sunspot in a counterclockwise direction (the square-framed zone in Figure \ref{fig:tile}). Because this flux tube is not given any twist at the initial state, the observed sunspot rotations are presumed to be driven by the background turbulence beneath the solar surface.

\subsection*{Magnetic helicity injection and sunspot rotation}

\begin{figure}[ht]
\centering
\includegraphics[width=0.8\linewidth]{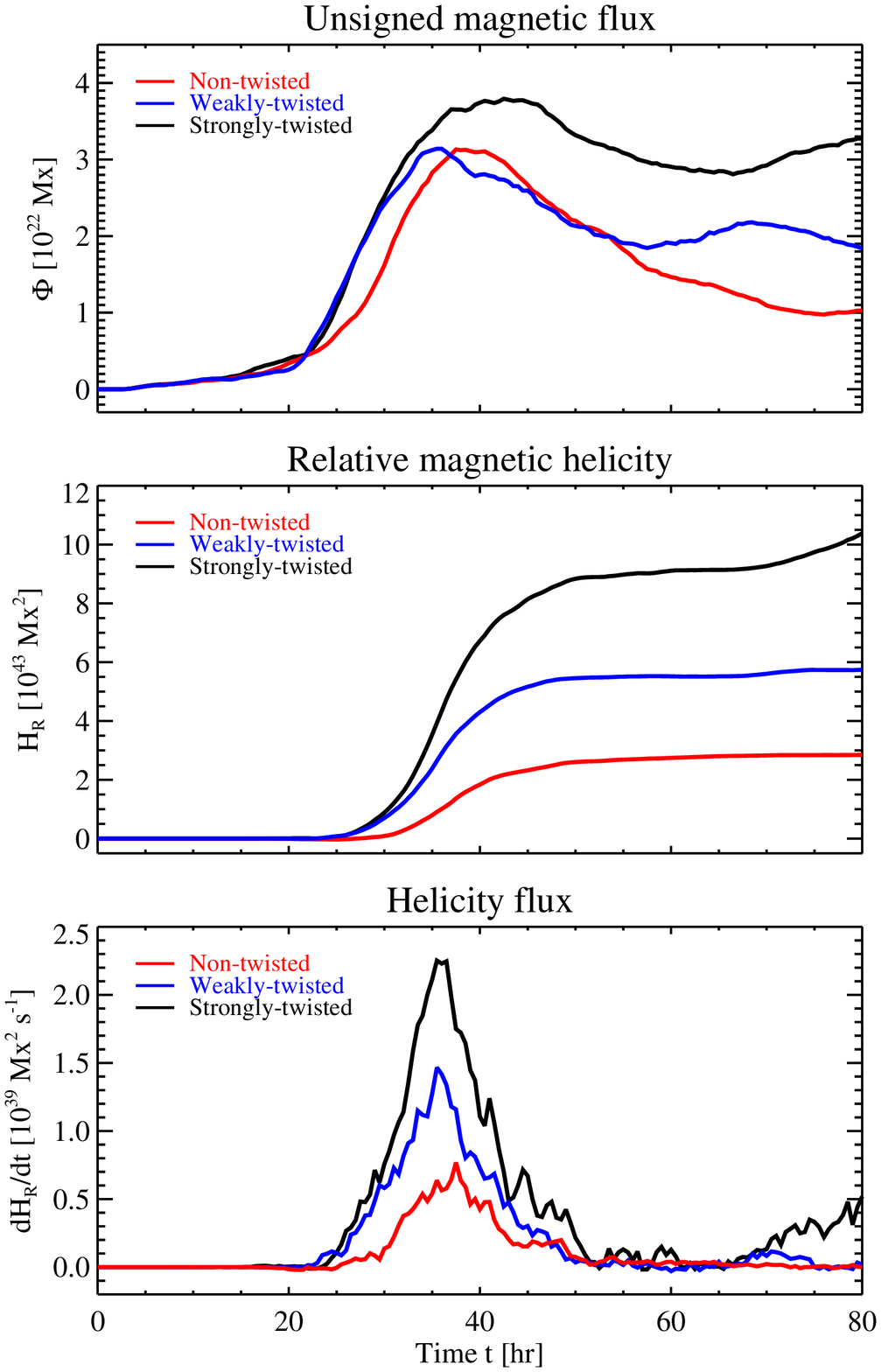}
\caption{Temporal evolutions of total unsigned magnetic flux ($\Phi$), relative magnetic helicity ($H_{\rm R}$) and helicity flux ($dH_{\rm R}/dt$) measured in the photosphere for the three cases.}
\label{fig:1d}
\end{figure}

Figure \ref{fig:1d} shows the temporal evolutions of injected magnetic helicity, $H_{\rm R}$, and helicity flux, $dH_{\rm R}/dt$, as compared with the total unsigned magnetic flux,
\begin{eqnarray}
  \Phi=\int |B_{z}|\, dS,
\end{eqnarray}
all measured in the photosphere for the three cases. The initial total magnetic flux in the flux tubes in the axial direction is of the order of $2\times 10^{22}\ {\rm Mx}$. Thus, if a flux tube emerges bodily as a whole to form a pair of sunspots, the total photospheric flux would be $4\times 10^{22}\ {\rm Mx}$. However, in all cases it is on the order of $3\times 10^{22}\ {\rm Mx}$, indicating that some fractions of the flux tubes remain below the photosphere. The total fluxes peaked around $t=40\ {\rm hr}$ and then gradually decreased as the sunspots decayed.

The middle panel of Figure \ref{fig:1d} illustrates that positive magnetic helicity is injected in the cases of strong (black line) and weak (blue line) twists. This is reasonable considering that these flux tubes were initially given right-handed twists and that the sunspots displayed clockwise rotations. What is noteworthy about this plot is that the positive helicity injection also occurs in the non-twisted case (red line). The accumulated helicity amounts as much as to $2.9\times 10^{43}\ {\rm Mx}^{2}$, which is about 20\% to 50\% of those in the twisted cases. This result reveals that a non-negligible amount of magnetic helicity can be injected even when a twist-free flux tube emerges in the convective zone with null net kinetic helicity. It should be noted that the magnetic helicity normalised by the square of the total photospheric flux, $H_{\rm R}/\max{(\Phi)}^{2}$, for the three cases is 0.029, 0.058 and 0.094 (peak values), which are comparable to the recent observations of flaring active regions\cite{2019A&A...628A..50M,2019ApJ...887...64T}. Therefore, the present simulations provide the reasonable reproduction of solar active regions. The bottom panel of Figure \ref{fig:1d} shows the evolution of helicity flux (see Equation (\ref{eq:dhdt})).

\begin{figure}[ht]
\centering
\includegraphics[width=\linewidth]{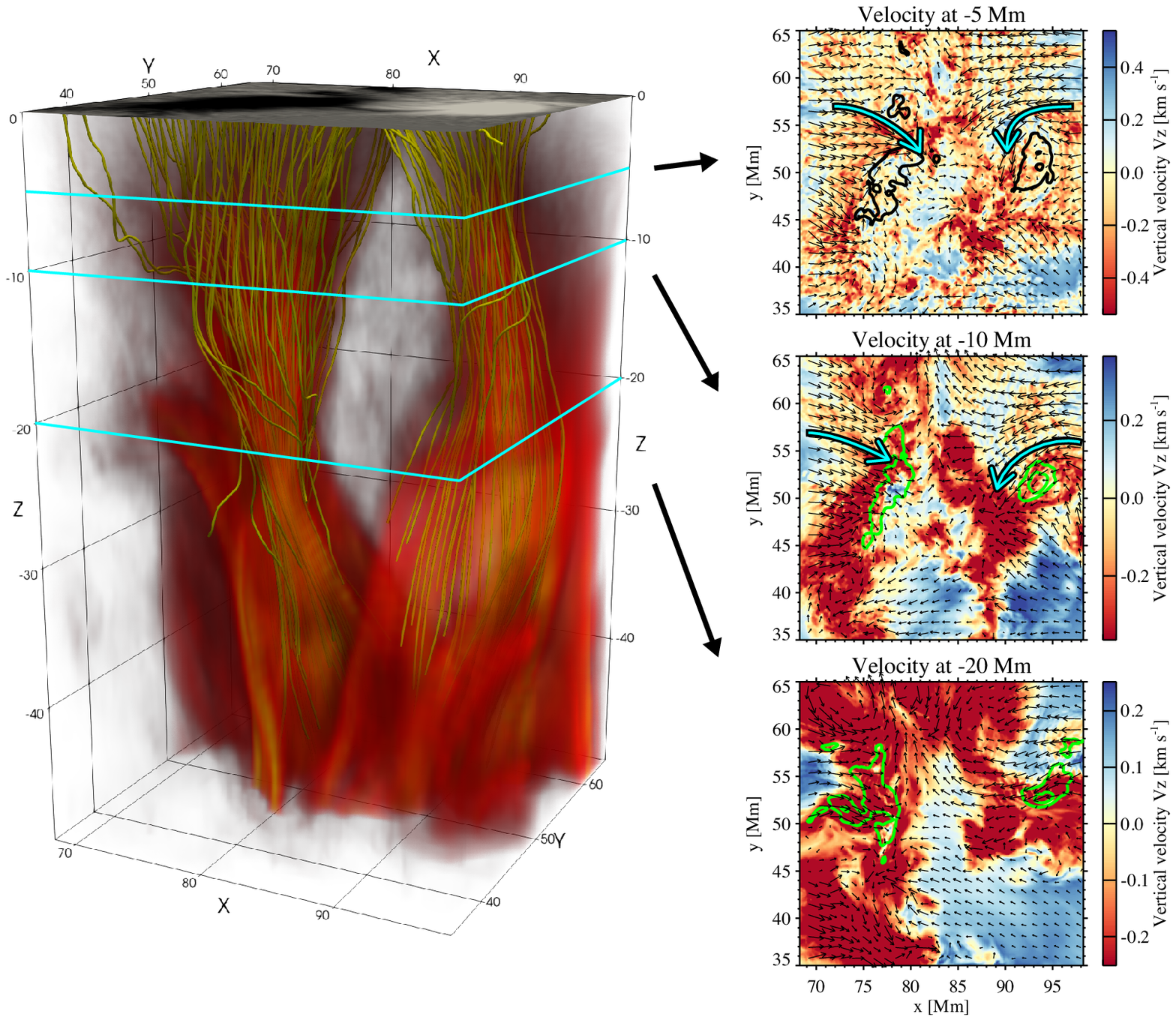}
\caption{3D view of the magnetic structures below the two rotating sunspots, whose field of view is indicated in Figure \ref{fig:tile}. The reddish volume rendering shows the total field strength ($|\mbox{\boldmath $B$}|$), while the yellow lines are the selected magnetic field lines. The 2D slice at the top presents the photospheric vertical field ($B_{z}$) at the $\tau=1$ surface, with the grayscale saturating at $\pm 2\ {\rm kG}$. The three panels on the right show the velocity fields ($V_{z}$ and $\mbox{\boldmath $V$}_{\rm h}$ vector) at $-5$, $-10$ and $-20\ {\rm Mm}$. Contours (black or green depending on the background) show the 50\% and 80\% levels of the maximum total field strength ($|\mbox{\boldmath $B$}|$) of the entire horizontal (i.e. $98.3\ {\rm Mm} \times 98.3\ {\rm Mm}$) plane at each depth, representing the two magnetic pillars. Cyan arrows highlight the directions of the horizontal flows. All physical values are obtained by averaging the data sets over a period between $t=35.0\ {\rm hr}$ and 38.0 hr.
}
\label{fig:vortex}
\end{figure}

Then, what effect causes a finite positive helicity injection in the non-twisted case? Figure \ref{fig:vortex} shows the 3D rendering below the sunspot collision area (corresponding to the square-framed zone in Figure \ref{fig:tile}) and the plots of velocity fields at three different depths. We average the data over the period from $t=35\ {\rm hr}$ to 38 hr, when the helicity flux is peaked (bottom panel of Figure \ref{fig:1d}). Below the positive and negative sunspots, a pair of vertical magnetic fluxes (represented by yellow magnetic field lines) extend downward into the deep convection zone. The velocity plots reveal that the two magnetic pillars (indicated by contours) reside in the strong downflowing plumes, to which surrounding plasmas stream in, leaving local vortices (highlighted by cyan arrows). These structures are created because some portions of the initially horizontal flux tube are dragged into the strong downflow plumes and become vertical magnetic concentrations\cite{2017ApJ...846..149C,2020MNRAS.494.2523H}. At the same time, the plumes also drive the surrounding plasmas to flow into them, accompanied by the vortices. The directions of the vortices agree with those of the sunspot rotations in the photosphere, i.e., clockwise (counterclockwise) for the negative (positive) sunspot. This indicates that the local vortices streaming into the downflow plumes spin the magnetic pillars and drive the sunspot rotations in the photosphere. If the sunspot rotation is because of the release of magnetic twists, the two magnetic pillars (sunspots in the photosphere) should rotate in the same direction\cite{2009ApJ...697.1529F,2015A&A...582A..76S}. However, this is not the case. Thus, even if the flux tube is initially endowed with no magnetic twist, it is possible that the surrounding turbulent flow, in which the local vortices reside, exerts a spinning effect on the flux tube and, accordingly, the sunspot rotation occurs in the photosphere, leading to the significant injection of magnetic helicity into the upper atmosphere.

\subsection*{Flare productivity}

\begin{figure}[ht]
\centering
\includegraphics[width=0.8\linewidth]{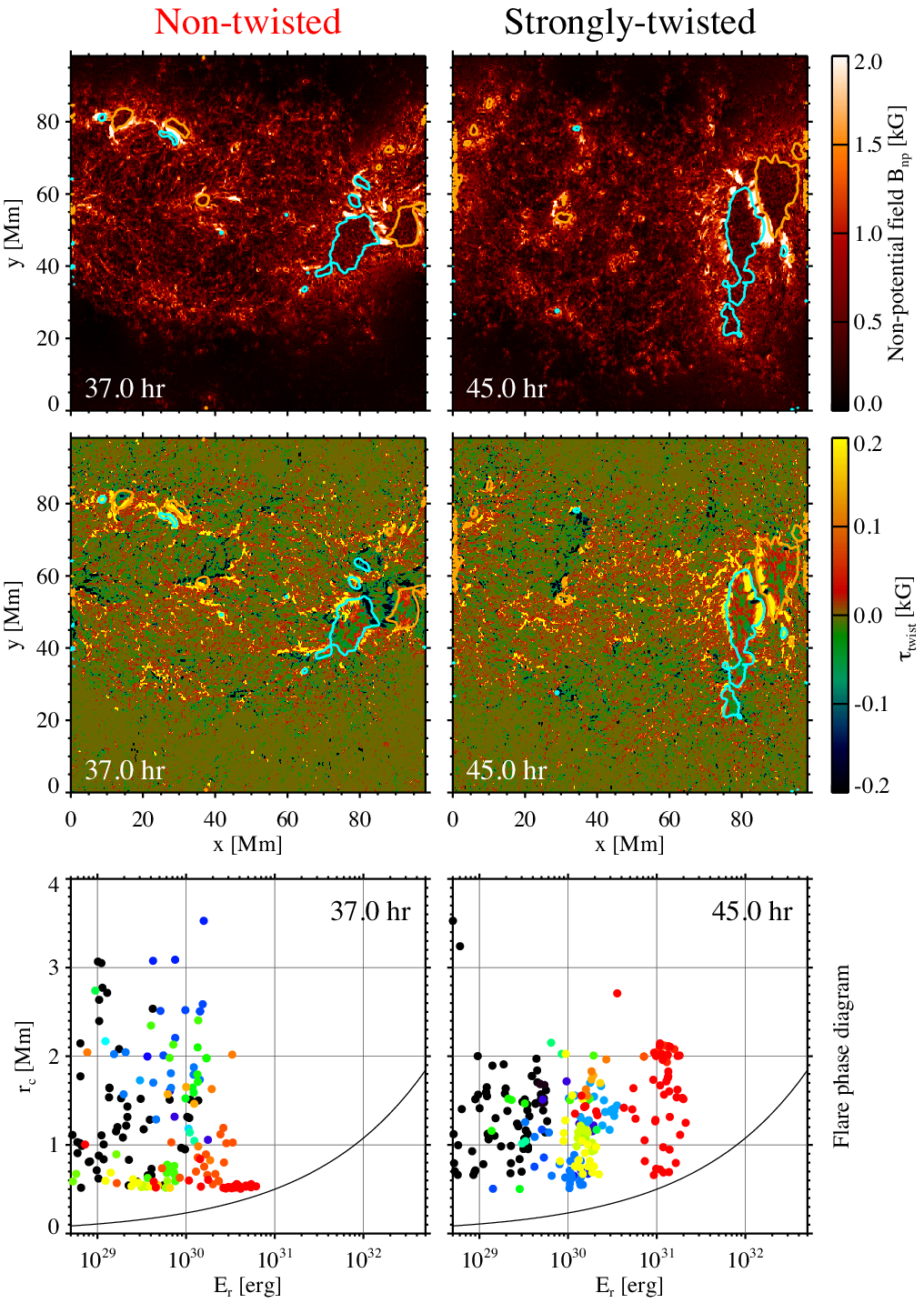}
\caption{(Top) Non-potential magnetic field ($B_{\rm np}=|\mbox{\boldmath $B$}-\mbox{\boldmath $B$}_{\rm p}|$) for the non-twisted and strongly-twisted cases. The cyan and orange lines indicate the smoothed contours of $B_{z}=-2\ {\rm kG}$ and 2 kG, respectively. (Middle) Magnetic twist flux density ($\tau_{\rm twist}=T_{\rm W} |B_{z}|$). Contours are identical to those in the top panels. (Bottom: flare phase diagram) Critical radius of the circular reconnection region required to satisfy the double-arc instability (DAI) condition, $r_{\rm c}$, and the magnetic energy that can be released, $E_{\rm r}$, for all points on the polarity inversion lines in the identified high free-energy regions (HiFERs). The colours denote the different HiFER patches, with the reddish (blueish) colours for the larger (smaller) patches. Observations show that X2-class flares or larger are likely if there are points in the area of $E_{\rm r}>4\times 10^{31}\ {\rm erg}$ and $r_{\rm c}<1\ {\rm Mm}$\cite{2020Sci...369..587K}. The black curve represents the self-similar scaling of $r_{\rm c}\propto E_{\rm r}^{1/3}$.
}
\label{fig:hifer}
\end{figure}

One of the most promising methodologies that have been suggested to predict the flare eruptions is to analyse the spatial distribution of high free-energy regions (HiFERs), where the non-potential magnetic field, $B_{\rm np}=|\mbox{\boldmath $B$}-\mbox{\boldmath $B$}_{\rm p}|$, exceeds the critical value of 1 kG, and examine the occurrence of the double-arc instability (DAI\cite{2017ApJ...843..101I}) for each HiFER patch near the polarity inversion line\cite{2020Sci...369..587K}. Top and middle panels of Figure \ref{fig:hifer} show the distributions of $B_{\rm np}$ and the magnetic twist flux density $\tau_{\rm twist}$ at the representative timings for the non-twisted and strongly-twisted cases. The bottom panels are the flare phase diagram, which is the scatterplot of the minimum critical radius of the circular reconnection region required to satisfy the DAI condition, $r_{\rm c}$, versus the magnetic energy that can be released, $E_{\rm r}$, for the identified HiFERs. Observations show that large X-class flares likely occur when there is a HiFER with $r_{\rm c}<1\ {\rm Mm}$ and $E_{\rm r}>4\times 10^{31}\ {\rm erg}$ on the flare phase diagram\cite{2020Sci...369..587K}.

In the strongly-twisted case, the HiFER patches ($B_{\rm np}>1\ {\rm kG}$) are coherent and the positive twist is distributed at the polarity inversion line between the two major sunspots, which is in agreement with the analysis results in the previous subsections. In the flare phase diagram, several locations are very close to the X-flare criticality. Therefore, we can predict that M-class flares are possible with a slight chance of X flares for the strong twist case. In the non-twisted case, however, although there are regions where $B_{\rm np}$ exceeds 2 kG, the HiFERs are more fragmented. The twists around the polarity inversion lines are a mixture of positive and negative signs, suggesting that the twists of both signs can be generated, probably randomly, because of the nature of turbulent convection, when a twist-less magnetic flux emerges. No HiFERs satisfy the X-flare threshold in the flare phase diagram, but lower-class (e.g. C-class) flares may still occur.

\section*{Discussion}

We investigate the physical mechanisms that provide magnetic helicity into the solar corona, which is critically important in understanding the occurrence of flare eruptions, by comparing the flux emergence simulations with and without magnetic twists under the existence of turbulent thermal convection. It is widely believed that the emergence of twisted flux tubes and associated sunspot rotations supply helicity into the active region corona. However, this is not trivial because of the lack of optical observations below the photosphere, which is the primary motivation of this study. We summarise the key results as follows.

\begin{enumerate}
\item A non-twisted flux tube can reach the photosphere and develop an active region. This is contrary to the previous simulation results without thermal convection during the emergence, a rising untwisted flux tube splits into two vortex rolls and quickly loses its identity because of the counteracting aerodynamic drag\cite{2021LRSP...18....5F}. With the support of convection, a non-twisted flux tube may complete its journey to the solar surface.

\item The injected magnetic helicity is non-zero even in the case of no magnetic twist. In fact, the amount of injected helicity was quite large, reaching about 50\% as in the twisted cases. Our analysis reveals that in the convection zone below the developed sunspots, inflows into the strong downwelling plumes produce local vortex motions, which spin the vertically standing magnetic flux tubes that extend along the plumes. As a result, the sunspots in the photosphere show rotations in the directions determined by the subsurface vortices, and a considerable amount of helicity injection into the upper atmosphere occurs.

\item The instability analysis suggests that the generated non-potential magnetic field in the non-twisted case can produce eruptions, albeit weaker than the twisted cases. This indicates that the ``twist'' observed in a variety of forms in the flare-productive active regions of the Sun is not only supplied by the magnetic ``twist'' of the emerging flux and the associated sunspot rotations, as previously believed, but also includes a non-negligible fraction of the ``twist'' brought by the background turbulent convection.
\end{enumerate}

In the present simulations, although the net kinetic helicity of the initial background convection was infinitesimally small, the resultant helicity injection in the photosphere is comparable to that provided by the emergence of twisted flux tubes. Considering the nature of the turbulence where we do not consider the solar rotation, the directions of the rotations of the local vortices that contributed to the spinning of the flux tubes may probably be determined by chance. Because the downflow plumes accompanied by the vortices are a coherent structure owing to the large time scale in the deeper convection zone, although the directions of the vortices are randomly determined, the sign of helicity injection may remain the same (positive in the present runs) over time. This indicates that the sign of helicity injection in simulations with different background turbulence can be negative, which will be examined in the future studies.

Asymmetry in sunspot rotation is reported in bipolar active regions\cite{2007A&A...468.1083Y,2008MNRAS.391.1887Y}, and this could be due to the difference in vortex motions in the convection zone. Also, a considerable fraction of solar active regions do not obey the hemispheric helicity sign rule, with the fraction of violators being 20--40\% \cite{2014SSRv..186..285P,2020ApJ...904....6P}. This randomness may be the result of the stochastic imposing of magnetic helicity by the background turbulence to the twist of the emerging flux that is determined by the solar dynamo\cite{2016Sci...351.1427H}.

\section*{Methods}

In this study, we use the radiative MHD code {\it R2D2}\cite{2019SciA....5.2307H}, which realistically simulates the thermal convection over the entire solar convection zone, to model the emergence of magnetic flux tubes and the spontaneous sunspot formation\cite{2019ApJ...886L..21T,2020MNRAS.494.2523H,2020MNRAS.498.2925H,2022MNRAS.517.2775K}. This code computes the 3D MHD equations with realistic radiation transfer and equation of state, and by implementing RSST\cite{2005ApJ...622.1320R,2012A&A...539A..30H,2015ApJ...798...51H,2019A&A...622A.157I}, it effectively suppresses the high sound speed in the deep solar interior and relaxes the Courant-Friedrichs-Lewy condition.

We use a rectangular Cartesian box as the computational domain, spanning over $0\leq x\leq 98.3 \ {\rm Mm}$, $0\leq y\leq 98.3\ {\rm Mm}$, and $-201\ {\rm Mm}\leq z\leq 676\ {\rm km}$, resolved by a $1024\times 1024\times 384$ grid. The grid spacing for the horizontal directions was $\Delta x=\Delta y=96\ {\rm km}$ (uniform), while that for the vertical direction was uniform at $\Delta z=48\ {\rm km}$ from the top boundary to $z=-5.6\ {\rm Mm}$, which linearly increased to the bottom boundary up to $\Delta z=1486\ {\rm km}$. The horizontal boundaries assumed the periodic boundary condition, while the magnetic field at the top boundary was connected to the potential field. The initial background convection was the same as that in\cite{2020MNRAS.498.2925H}, in which a magnetic flux tube emerges to successfully produce a bipolar sunspot group. The net kinetic helicity is negligibly small because the calculation box did not consider the solar rotation. The normalised net kinetic helicity at $t=0$, when the flux tube was embedded, was;
\begin{eqnarray}
  \frac{\displaystyle\int_{z< 0} \mbox{\boldmath $V$}\cdot(\nabla\times \mbox{\boldmath $V$})\, dV}{\displaystyle\int_{z<0}|\mbox{\boldmath $V$}\cdot(\nabla\times \mbox{\boldmath $V$})|\, dV}
  =0.00523\%.
\end{eqnarray}

The magnetic flux tube was embedded at a depth of 22 Mm. Unlike the previous {\it R2D2} simulations, which used the force-free magnetic flux tubes, we adopt commonly used Gaussian-type flux tubes, where the longitudinal and azimuthal components of the magnetic field are given by
\begin{eqnarray}
  B_{x}(r)=B_{\rm tb}\exp{\left(-\frac{r^{2}}{R_{\rm tb}^{2}}\right)}
\end{eqnarray}
and
\begin{eqnarray}
  B_{\phi}(r)=qrB_{x}(r),
\end{eqnarray}
where $B_{\rm tb}$, $r$, $R_{\rm tb}$ and $q$ are the axial field strength, radial distance from the tube axis, typical radius and twist intensity, respectively. We calculate three cases with varying $q$ (Table \ref{tab:case}). Here, $B_{\rm tb}$ was also varied so that the magnetic energies of the flux tubes, $E_{\rm mag}(=\int B^{2}/(8\pi)\, dV)$, are the same in all three cases. The total axial magnetic flux $\Phi_{x}\,(=\int B_{x}\,dS)$ is of the order of $2\times 10^{22}\ {\rm Mx}$. To achieve the mechanical balance with no magnetic buoyancy, we adjust the internal pressure and density by reducing the entropy by $\frac{B^2}{8\pi}/\left(\frac{\partial p}{\partial s}\right)_{\rho}$. Therefore, the flux tubes started emergence only because of advection exerted by the background turbulence. In all three cases, we set $q$ to be 0 or below the critical value for the kink instability $q_{\rm cr}\,(=1/R_{\rm tb})$\cite{1996ApJ...469..954L}, where $q>0$ indicates that the tube has a right-handed twist.

\begin{table}[ht]
\centering
\begin{tabular}{|l|l|l|l|l|l|}
\hline
Case & $R_{\rm tb}$ & $q$ & $B_{\rm tb}$ & $E_{\rm mag}$ & $\Phi_{x}$ \\
\hline
Non-twisted & 8 Mm & 0 & 12.2 kG & $5.9\times 10^{34}\ {\rm erg}$ & $2.4\times 10^{22}\ {\rm Mx}$\\
\hline
Weakly-twisted & 8 Mm & $q_{\rm cr}/4$ & 12.1 kG & $5.9\times 10^{34}\ {\rm erg}$ & $2.4\times 10^{22}\ {\rm Mx}$ \\
\hline
Strongly-twisted & 8 Mm & $q_{\rm cr}/2$ & 11.5 kG & $5.9\times 10^{34}\ {\rm erg}$ & $2.3\times 10^{22}\ {\rm Mx}$ \\
\hline
\end{tabular}
\caption{\label{tab:case}Summary of the simulation cases. The critical twist for the kink instability corresponds to $q_{\rm cr}=1/R_{\rm tb}=0.125\ {\rm Mm}^{-1}$.}
\end{table}

We measure several physical quantities including the relative magnetic helicity ($H_{\rm R}$), helicity flux ($dH_{\rm R}/dt$) and total unsigned magnetic flux ($\Phi$) at the photospheric surface, here defined as where the optical depth is unity ($\tau=1$). For measuring the helicity, there is a degree of freedom in choosing the vector potential $\mbox{\boldmath $A$}_{\rm p}$ and, in this study, we select vector potentials satisfying $\mbox{\boldmath $A$}_{\rm p}\cdot\hat{\mbox{\boldmath $z$}}=0$ and calculate it using the method by\cite{2005A&A...439.1191P},
\begin{eqnarray}
\mbox{\boldmath $A$}_{\rm p}=\frac{1}{2\pi}\hat{\mbox{\boldmath $z$}}\times
\int_{S'}B_{z}(\mbox{\boldmath $x$}')\frac{\mbox{\boldmath $r$}}{r^{2}}\, dS',
\end{eqnarray}
where $\mbox{\boldmath $r$}=\mbox{\boldmath $x$}-\mbox{\boldmath $x$}'$. In this process, the grid number was reduced from $1024\times 1024$ to $256\times 256$ to accelerate the computation speed. The helicity flux was then calculated using Equation (\ref{eq:dhdt}), and the injected magnetic helicity over the course of flux emergence (Equation (\ref{eq:hr})) was obtained by integrating the helicity flux over time.

To investigate the flare productivity of the generated active regions, the instability analysis is performed based on the DAI theory, which states that a flux rope becomes unstable if it is sufficiently twisted and has a sufficient amount of magnetic flux against the overlying confinement field\cite{2017ApJ...843..101I}. This instability is characterised by the parameter
\begin{eqnarray}
  \kappa=\left|\frac{\displaystyle\int_{\rm rec} T_{\rm W}\, d\Phi}{\displaystyle\Phi_{\rm over}}\right|,
\end{eqnarray}
or equivalently,
\begin{eqnarray}
  \kappa=\left|\frac{\displaystyle\int_{S_{\rm rec}} \tau_{\rm twist}\, dS}{\displaystyle\Phi_{\rm over}}\right|,
\end{eqnarray}
where $T_{\rm W}$ is the amount of twist integrated over each magnetic field line in a flux rope, $S_{\rm rec}$ is the footpoint area of the magnetic field lines that reconnect to form the flux rope, $\Phi_{\rm over}$ is the magnetic flux of the overlying field, and $\tau_{\rm twist}=T_{\rm W} |B_{z}|$. In DAI, $\kappa$ is usually larger for larger $S_{\rm rec}$; therefore, there is a critical value $S_{\rm c}$ at which the instability occurs: $\kappa>\kappa_{0}\sim 0.1$. It is shown that the ratio of the magnetic helicity of current-carrying magnetic field ($|H_{\rm j}|$) to the total relative helicity ($|H_{\rm R}|$) well discriminates whether a flare event becomes eruptive or not\cite{2017A&A...601A.125P}. While the $\kappa$ parameter is the critical parameter which determines the onset of DAI (i.e., whether a flare occur), the helicity ratio $|H_{\rm j}|/|H_{\rm R}|$ may be capable of distinguishing the eruptivity when the flare occurs.

In the $\kappa$-scheme\cite{2020Sci...369..587K}, the coronal magnetic field is first calculated by the non-linear force-free field extrapolation based on the vector magnetic field of the photosphere. For the computational resources, the grid number was reduced from $1024\times 1024$ to $512\times 256$ in this process. Then, HiFERs are identified as the regions where the non-potential field $B_{\rm np}(=|\mbox{\boldmath $B$}-\mbox{\boldmath $B$}_{\rm p}|)$ exceeds the threshold value of 1 kG, which is based on the observational result that $B_{\rm np}=1$ kG sufficiently encompasses the the distribution of non-potential magnetic fields that drive large flares\cite{2020Sci...369..587K}. For each HiFER, the critical area $S_{\rm c}$ is measured as the minimum circular area that satisfies the DAI condition ($\kappa>\kappa_{0}$), and $r_{\rm c}$ is obtained as the radius of $S_{\rm c}$, i.e., $S_{\rm c}=\pi r_{\rm c}^{2}$. The releasable energy for each HiFER is estimated as
\begin{eqnarray}
  E_{\rm r}=\frac{S_{\rm r}^{1/2}}{8\pi} \int_{S_{\rm r}} B_{\rm np}^{2}\, dS,
\end{eqnarray}
where $S_{\rm r}$ is the area of the footpoint of the magnetic flux that pass over the circular area $S_{\rm c}$.

\section*{Data Availability}

The data are available from the corresponding author upon reasonable request.

\bibliography{sample}

\section*{Acknowledgements}

The initial convection data were obtained using the Supercomputer Fugaku provided by the RIKEN Center for Computational Science. Numerical computations were in part carried out on Cray XC50 at Center for Computational Astrophysics, National Astronomical Observatory of Japan. This work was supported by JSPS KAKENHI Grant Nos. JP20KK0072 (PI: S. Toriumi), JP20K14510 (PI: H. Hotta), JP21H01124 (PI: T. Yokoyama), JP21H04492 (PI: K. Kusano), JP21H04497 (PI: H. Miyahara), and by MEXT as a Program for Promoting Researches on the Supercomputer Fugaku.

\section*{Author contributions statement}

S.T. conducted the simulations, analysed the results, and wrote the manuscript. H.H. developed the simulation code and prepared the initial convection data. K.K. conducted the $\kappa$-scheme analysis. All authors reviewed the manuscript.

\section*{Additional information}

\textbf{Correspondence} and requests for materials should be addressed to S.T.

\section*{Competing interests}

The authors declare no competing interests.

\end{document}